\documentclass[final,3p,twocolumn]{elsarticle}

\usepackage{graphics}
\usepackage{epstopdf}
\usepackage{epsfig}

\usepackage{amssymb}

\usepackage{lineno}

\journal{Astroparticle Physics}

\begin{document}

\begin{frontmatter}



\title{Rejection of  Surface Background in Thermal Detectors}
\author[a]{Carlo Bucci\corref{io}}
\cortext[io]{Corresponding author.}
\ead{carlo.bucci@lngs.infn.it}
\author[a]{Paolo Gorla}
\author[b]{Wolfgang Seidel}
\address[a]{INFN - Laboratori Nazionali del Gran Sasso, I-67010 Assergi (L'Aquila), Italy}
\address[b]{Max-Planck-Institut f\"ur Physik, F\"ohringer Ring 6, D-80805 Munich, Germany}
\begin{abstract}
Thermal detectors have recently achieved a leading position in the fields of neutrinoless double beta decay and dark matter searches thanks to excellent energy resolution and numerous absorber material choices. These detectors operate at temperatures of a few mK and do not have dead layers. Contaminations on the surface of faced materials have thus to be taken into account as sources of background. We propose a scintillation-based approach for rejecting surface background and discuss the innovative application of this technique in non-scintillating bolometric detectors.
\end{abstract}


\end{frontmatter}

\section{Introduction}

In recent years large mass ($\sim$1 kg) thermal detectors have achieved excellent energy resolution, of the order of a few per mil. Due to the vast choice of absorber materials possible, which is limited only by the heat capacity at the operating temperature, thermal detectors are presently successfully employed in many dark matter (DM) \cite{CRESST,CDMS,Edelweiss,Rosebud} and neutrinoless double beta decay (0$\nu$DBD) \cite{Cuoricino,CUORE} searches.

In such types of experiments, which look for extremely rare events, control of backgrounds is the main concern. An intrinsic downside to bolometers is that they lack the capability to distinguish between different types of interacting particles (e.g., electrons vs. alpha particles vs. nuclear recoils).
In recent years many DM experiments have overcome this limitation by developing bolometers which provide light yield or ionization information in addition to the heat channel.
Moreover, thermal detectors operate in vacuum and have a fully sensitive volume, which permits efficient application of anti-coincidence techniques for detector arrays. However, the absorbers are also exposed to events coming from surfaces of materials facing the detectors, which cannot be distinguished from bulk events.

The CDMS and EDELWEISS experiments use germanium detectors and the particle identification is performed by simultaneously reading ionization and thermal signals. Surface events can give rise to an incomplete charge collection, simulating nuclear recoils. For this reason strategies based on interleaved electrodes \cite{Edelweiss_surface} or timing of pulses \cite{CDMS, CDMS_surface} have been developed to identify surface events.

The CRESST experiment employs scintillating crystals which acquire, in addition to the thermal signal, the scintillation light emitted inside the absorber by the particle interaction. The light is detected by specially developed thermal detectors. In order to discriminate surface events, any passive item surrounding the detector is also covered with scintillating material.

We believe that the same technique could successfully be adopted in experiments using purely thermal detectors where the absorber does not scintillate. We discuss the potential of this technique when applied to an experiment using arrays of pure (non-scintillating) bolometric detectors.

\section{Pure bolometric detectors: the CUORE case}

CUORE (Cryogenic Underground Observatory for Rare Events)  \cite{CUOREproposal,CUORE} is an upcoming experiment searching for the 0$\nu$DBD of $^{130}$Te. It  will consist of an array of 988 TeO$_2$ thermal detectors operating at $\sim$10 mK with a total mass of $\sim$750 kg. Nuclear Transmutation Doped (NTD) germanium chips  are used as temperature sensors. CUORE is the largest example of pure bolometric experiment. 
Its background will be dominated by radioactive contaminations on the surfaces of the materials facing the detectors, and many methods have been devised in order to get rid of this contribution. A background of less than 0.01 counts/keV/kg/y in the energy region of interest is required in order to reach the sensitivity necessary to probe the inverse hierarchy for the effective Majorana neutrino mass.

Cuoricino \cite{Cuoricino}, a 40 kg-scale CUORE prototype, has been useful for understanding background sources. The energy spectrum measured in Cuoricino indicates that the CUORE background in the 0$\nu$DBD region will be dominated by degraded alpha particles emitted from the passive materials that face the crystal absorbers, such as the copper structures holding the detectors. 
These alphas are emitted by radioactive nuclei implanted in the copper due to manufacturing or environmental contaminations. An alpha particle emitted by the contaminated surface can escape the material before releasing all its energy. As a result, the spectrum of escaped alphas is a continuum starting at the nominal energy of the alpha (the daughter nucleus is emitted in the opposite direction and stays on the copper) and decreasing to zero. Therefore a copper surface facing a bolometer can generate a continuous background in the  0$\nu$DBD region \cite{bolometer_background,alpha_background}.
The level of surface contamination on the copper produces a few alpha decays/cm$^2$/year. This value is very low but is still too high to reach the target background level. 

\section{Experimental strategies against surface background}
\label{Pres_Strat}

In this section we describe several strategies conceived to reduce or discriminate alpha particles interacting on the crystal surfaces. We first describe the cleaning and passive-shielding strategies currently adopted in CUORE, then present some active rejection techniques that have been developed as possible alternative solutions, and finally introduce some of the ideas proposed for a future upgrade of CUORE.

The first obvious strategy for reducing surface contamination is to carefully clean all parts close to the detectors. Several cleaning methods have been tried, independently and in series, such as tumbling, chemical etching,  electropolishing and plasma cleaning. Despite all these efforts the measured background has not been reduced enough.

A second approach is the so-called passive shielding of the detector. This method consists of covering the surface of the materials facing the detector with very pure thin polyethylene foils. If the polyethylene multilayer is thick enough, it will stop alphas emitted from contaminated surfaces. This method is limited by the difficulty of full coverage (due to the complexity of the copper frames geometry) and by the cleanliness of the polyethylene foil (which itself can have surface contaminations). 

Due to the difficulties encountered in both the cleaning process and the passive shielding and to the incomplete removal of contaminations, several approaches based on active background reduction have been proposed. 
Since the 0$\nu$DBD signal consists of the simultaneous emission of two electrons which share the total energy of the reaction, the discrimination technique has to be able to distinguish between a bulk beta event and a surface alpha event next to the reaction Q-value, which for $^{130}$Te is roughly 2.5 MeV. 

As a consequence active approaches are based either on discrimination between surface and bulk events or  on the identification of the particle type (alpha vs. beta). In the following we describe first two attempts of rejecting surface events and them some proposal for  discriminating alpha/beta events in the detectors. 

The Surface Sensitive Bolometer (SSB) approach, a method based on the use of thin thermal detectors glued over the  TeO$_2$ surfaces has been proposed and tested  \cite{SSB}. By looking at the thermal pulse shapes of events releasing part of their energy directly in the thin detectors, discrimination of surface interactions is possible. The results have been encouraging and the discrimination capability has been demonstrated. Nevertheless the efficiency was limited by the difficulty in the realization of the full coverage of the crystal. This problem together with the complexity of the detector, which needs either multiple readout or very sensitive pulse shape discrimination, overwhelmed the possible advantages.

A second possibility that has been investigated is the use of NbSi thin films deposited on the TeO$_2$ crystal surfaces \cite{NbSi}. These films are capable of detecting out-of-equilibrium phonons, so when a particle releases energy in their proximity these films produce a higher and faster pulse. Discrimination between surface events should in principle be possible by equipping every face of a detector with a NbSi film.
The drawbacks of this technique are the non-negligible heat capacity of NbSi, which  could affect the detector performances, and the use of many readout channels, that makes the detector more elaborate. Also in this case a large scale application seems too complex. 

The idea of alpha/beta discrimination inspired a dual readout approach based on the simultaneous measurement of the thermal signal and scintillation light \cite{scint1,scint2,scint3}. 
By using this technique it is possible to recognize the nature of the particle releasing energy in the main thermal detector by observing different light yields. 
A crucial aspect of this approach is the detection of scintillation light at mK temperatures that is not trivial and implies the development of dedicated large area thermal detectors optimized to sense scintillation photons. 
While this idea has already been demonstrated to be extremely powerful in various experiments \cite{CRESST,Rosebud}, the method is limited to materials that exhibit some scintillation emission and cannot be applied to all the possible relevant materials (e.g. TeO$_2$\footnote{Indeed there is a proposal \cite{Lucifer} to exchange in the future the detecting crystals of CUORE from TeO$_2$ to another scintillating crystal like ZnSe }).

\begin{figure}[h!] 
   \centering
   \includegraphics[width=7.5 cm]{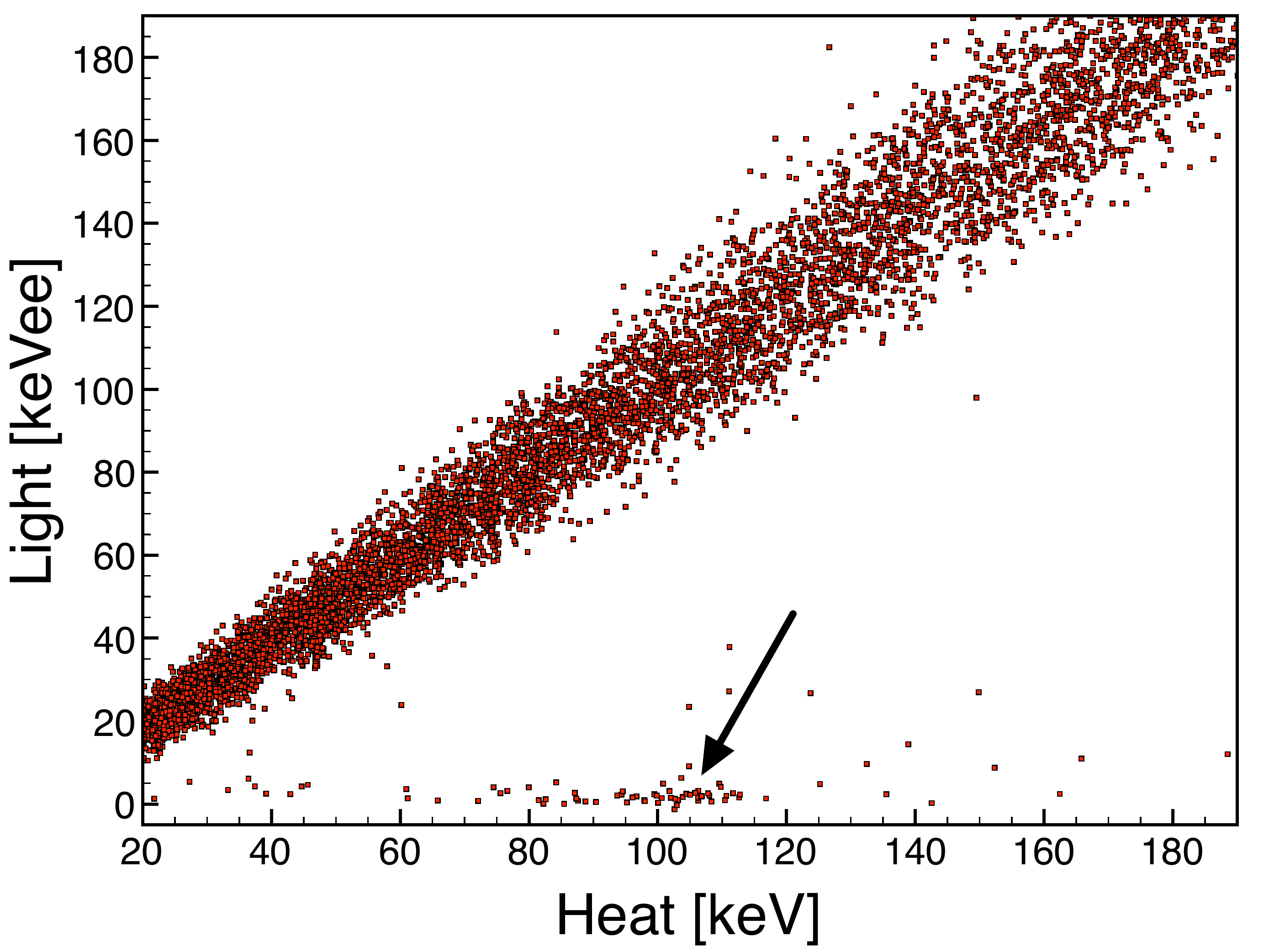}
   \includegraphics[width=7.5 cm]{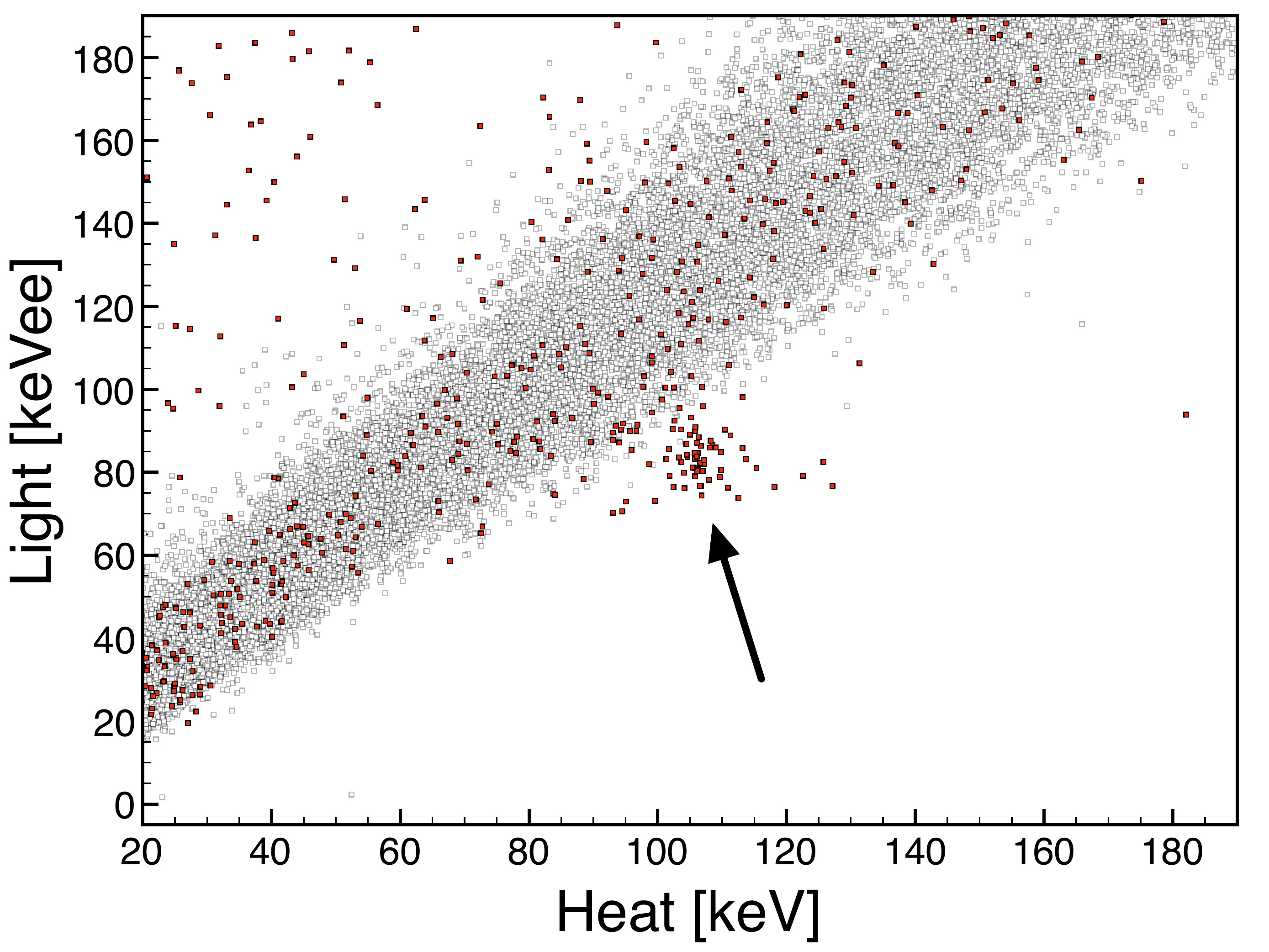}
   \caption{\em\small Scatter plot of light vs. heat of alpha decays on a CdWO$_{4}$ CRESST detector for a crystal encapsulated in a normal reflecting foil (upper picture) and in a scintillating reflecting foil (lower picture). Surface alpha decays  produce a clear spot around 100 keV on the heat channel (highlighted by black arrows). In the upper picture the measured light yield for these events is low and is due just to CaWO$_4$ scintillation properties. When the surrounding foil is also scintillating, the detected light is much larger because the alpha particle impacts on the foil; the events associated with scintillation of the foil were selected using pulse shape analysis on the light channel.}
   \label{f1}
\end{figure}

An attempt has been made to promote the low-temperature luminescence of TeO$_2$ by  doping \cite{TeO2nonscintilla}. Unfortunately, no scintillation light has been detected from the small samples in which doped elements were successfully incorporated.

Another possible approach, recently published  \cite{Cerenkov}, should allow the discrimination between alpha and beta particles by means of the reading of the Cerenkov light. This technique has the unquestionable advantage of relying on unaltered pure thermal detectors, but it requires the addition of an extremely sensitive light detector.
Indeed, the Cerenkov light emitted by a 0$\nu$DBD event of about 2.5 MeV is calculated to be approximately 125 photons which converts to roughly 350 eV. The fraction of energy released on the light detector will definitely be smaller due to losses from self-absorption (large crystals), total reflection (the refraction index of TeO$_2$ is about 2.4) and incomplete light collection.
Therefore it is very likely that the needed light detector must achieve a threshold significantly  lower than 100 eV, which is rather difficult to achieve at present.

\section{Proposed technique}

\begin{figure}[h] 
   \centering
   \includegraphics*[width=7.5 cm]{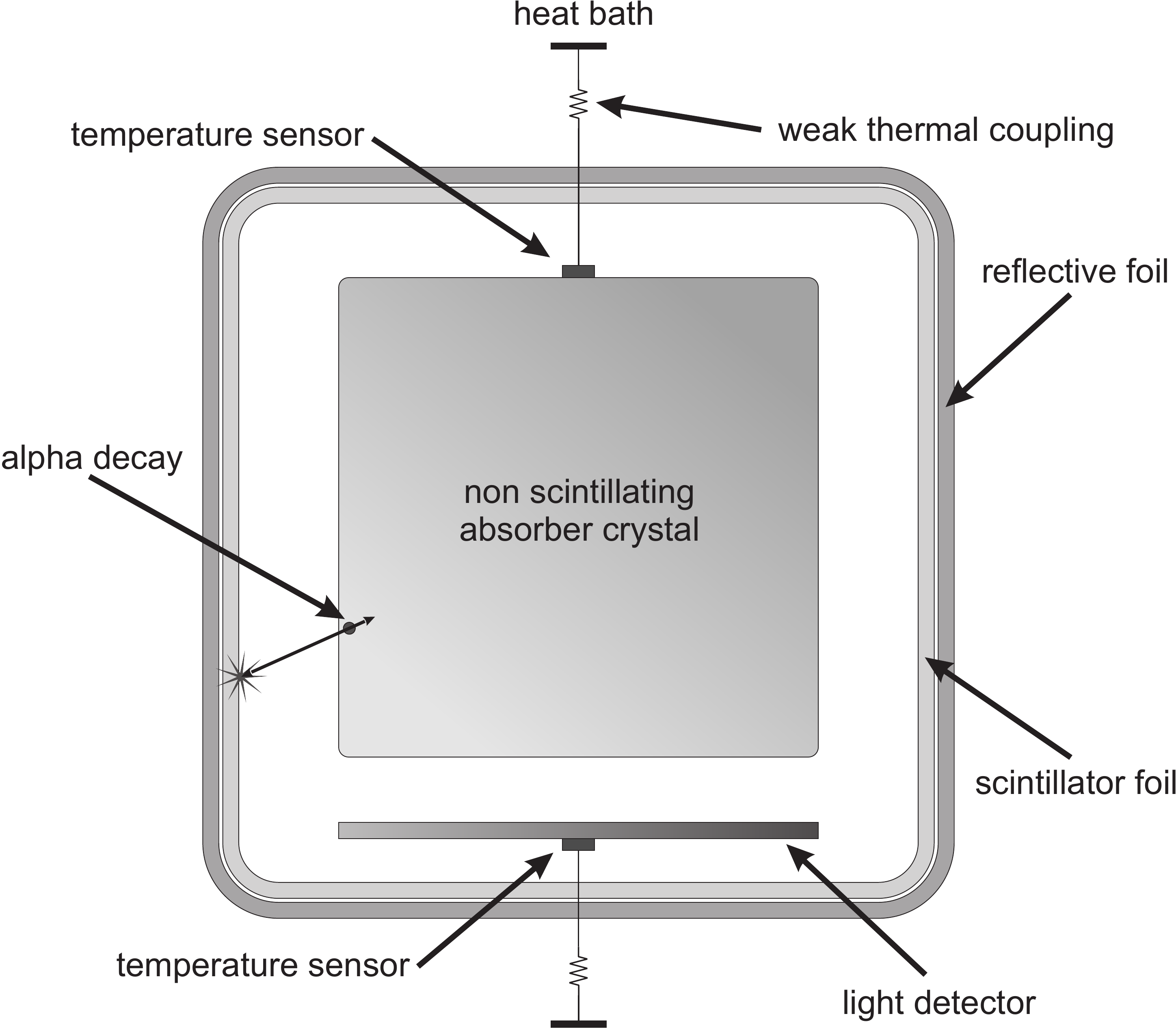}
   \caption{\em\small Diagram of the proposed method. When radiation escaping from the surface of a non-scintillating crystal strikes the reflecting and scintillating foil, light is emitted by the foil and detected by light detector. 
By analyzing light and heat events in coincidence, it is possible to tag surface events. The additional scintillator foil is advisable when a low-energy detection threshold is necessary.}
   \label{f2}
\end{figure}

We propose here to tag the events coming from surface contaminations by adopting an enhanced version of the technique successfully used by the CRESST experiment. Encapsulating a pure thermal detector with a scintillating foil (see Fig.~\ref{f2}), and adding a light detector, should make possible to identify degraded alpha particles. 

A surface alpha particle releasing part of its energy on the crystal and part on the scintillating foil can be tagged analyzing the coincidence signal of heat (in the absorber) and light (in the light detector) and rejected as background.

In the following we describe how this technique is applied in CRESST and which enhancements are necessary to apply it in CUORE.

The CaWO$_4$ detectors used in CRESST are surrounded by a highly reflective polymeric multilayer foil (VM2000/VM2002 from 3M) \cite{reflective_foil} in which the polyethylene naphthalate support was found to be a weak scintillator \cite{CRESST,reflective_foil_scintillates}.
This means that the foil, in addition to efficiently reflecting the photons emitted by particle interactions into  the absorber crystal, acts as a scintillator for particles impinging on the foil itself. This unique characteristic is very important in a DM experiment in which a critical background could be generated by alpha decays in which the alpha particle is emitted in the direction of the foil while the nuclear recoil impinges on the crystal.
The amount of scintillation light produced by the alpha particle in the reflecting foil is much larger than that generated by the recoiling nucleus in CaWO$_4$, offering a powerful tool for tagging these events. 

In Fig.~\ref{f1} the difference in the collected light between CaWO$_4$ detector with a normal reflector and with the scintillating foil is clearly visible. When a $^{210}$Po 5.3 MeV alpha particle hits the foil, the energy measured by the light detector is about 1 keV (deduced from the energy calibration of the light detector, performed with a $^{55}$Fe source).

In the case of the CUORE experiment, the goal is to discriminate alpha particles releasing energy in the TeO$_2$ crystals in the region of interest ($\sim$2.5 MeV). Since the less energetic alpha particles in the natural radioactive chains come from $^{232}$Th decay (Q-value of 4.01 MeV), the threshold of the light detector has to be low enough to tag alpha particles releasing an energy as low as 1.5 MeV on the scintillating foil.
Encapsulating the TeO$_2$ crystals in a scintillating reflecting foil should allow to tag the remaining alpha energy ($\sim$1.5 MeV) impinging on the foil. 
Unfortunately, plastic scintillators have an extremely non-linear response to alpha particles \cite{plastic_scintillator_at_lowE} and the light produced by an alpha particle of 1.5 MeV is about an order of magnitude smaller than at 5.3 MeV.  Detecting such a small amount of light represents a serious challenge for most of the currently used light detectors at mK temperature.

In order to detect low energy alpha particles it is therefore desirable to increase the amount of scintillation light.  
This can be simply achieved by superposing a thin plastic scintillator foil over the reflecting foil (see diagram in Fig.~\ref{f2}).
The scintillation light emitted by a commercial thin film plastic scintillator is approximately an order of magnitude larger with respect to the one measured in the 3M reflective foil. Moreover a conventional scintillator's light yield does not change appreciably when it is cooled to very low temperature, as has already been demonstrated in the study of polarized targets at mK temperature \cite{scintillating_foil_at_lowT}.
By providing a sufficiently thick scintillator (50 microns are enough to fully stop a 5 MeV alpha particle), the energy associated with the scintillation light of a 1.5 MeV alpha particle should be around 1 keV and thus easily detectable.
The small total  thickness of the two films ($< 200 \mu$m)  surrounding each detector guarantees that they will not significantly alter the coincidence detection of Compton events between neighboring detectors.  

The proposed  approach is effective also in the case of a beta surface contamination, because the smaller $dE/dx$ of electrons is counterbalanced by the much larger light yield they produce. The minimum energy released by an electron traversing a  50  $\mu$m scintillating foil is $\sim$50 keV. The light yield for a 50 keV electron is comparable to that emitted by a 2 MeV alpha particle and thus detectable by the light detector.
 
In summary,  this technique solves many of the problems which plague the other approaches described in Sec. \ref{Pres_Strat}.  The main advantage is that a discrimination tool is introduced without modifying the absorber that maintains its very powerful characteristics in terms of isotopic abundance and resolution. Moreover, in this case the light detector does not need to be extremely sensitive thanks to the high light yield of the plastic scintillating foil.  The cleanness of the material is not a crucial issue because the scintillating foil is active and radioactive decays inside it are easily tagged. 

Nevertheless, the necessity of adding a light detector, doubling the amount of read out channels, is a non-negligible downside.  This complexity can be substantially reduced using a single large-area light detector able to collect the scintillation light of multiple TeO$_2$ bolometers (e.g., in the current CUORE configuration, up to 8 bolometers could be read by a single light detector). 

As described above, this technique is effective both for surface alpha and beta particles, and its efficiency is related to the ability to fully cover the non-active components surrounding the detector with a scintillating material.

\section{Conclusions}

We have shown that the technique adopted to discriminate surface events in the CRESST experiment can be effectively employed by other experiments based on pure thermal detectors. 

For instance, with the successful application of this technique, a future upgrade of the CUORE experiment could strongly improve its background level.

\section*{Acknowledgments}
We would like to thank Tom Banks, Oliviero Cremonesi, Federica Petricca, and Franz Pr\"obst for useful discussions.






\begin{thebibliography}{00}


\bibitem{CRESST}
G.~Angloher et al.: \textit{Astroparticle Physics} \textbf{23} (2005) 325-339

\bibitem{CDMS}
Z.~Ahmed et al.: \textit{Physical Review Letters} \textbf{102}, (2009) 011301

\bibitem{Edelweiss}
A.~Benoit et al.: \textit{Physics Letters} \textbf{B616} (2005) 25-30

\bibitem{Rosebud}
S.~Cebrian et al.: \textit{Physics Letters} \textbf{B563} (2003) 48-52

\bibitem{Cuoricino}
C.~Arnaboldi et al.: \textit{Physical Review C} \textbf{78}, (2008) 035502

\bibitem{CUOREproposal}
R.~Ardito et al: \textit{arXiv}:hep-ex/0501010v1

\bibitem{CUORE}
C.~Arnaboldi et al.: \textit{Astroparticle Physics} \textbf{20} (2003) 91-110

\bibitem{Edelweiss_surface}
A.~Broniatowski et al., \textit{Journal of Low Temperature Physics} \textbf{151} (2008) 830-834 ;
A.~Broniatowski et al., \textit{Physics Letters} \textbf{B681} (2009) 305-309

\bibitem{CDMS_surface}
P.L.~Brink, et al.: \textit{Nuclear Instruments and Methods in Physics Research} \textbf{A559} (2006) 414-416

\bibitem{bolometer_background}
M.~Pavan et al.: \textit{European Physical Journal A} \textbf{36}, (2008) 159-166

\bibitem{alpha_background}
C.~Bucci et al.: \textit{European Physical Journal A} \textbf{41}, (2009) 155-168

\bibitem{Lucifer}
A.~Giuliani et al., LUCIFER: an Experimental Breakthrough in the Search for 
Neutrinoless Double Beta Decay, presented at BEYOND 2010, Cape Town, 
South Africa, February 1Ð6, 2010;

\bibitem{SSB}
L.~Foggetta et al.:
\textit{Applied Physics Letters}, \textbf{86}, Issue 13, (2005) 134106

\bibitem{NbSi}
C.~Nones et. al: \textit{Journal of Low Temperature Physics} \textbf{151}, (2008) 871-876

\bibitem{scint1}
L. Gonzalez-Mestres and D. Perret-Gallix, \textit{Nuclear Instruments and Methods in Physics Research} \textbf{A 279}, (1989) 382Ð387

\bibitem{scint2}
A. Alessandrello et al.,  \textit{Physics Letters B} \textbf{420} (1998), 109Ð113

\bibitem{scint3}
P.~Meunier et al., \textit{Applied Physics Letters} \textbf{75} (1999) 1335-1337

\bibitem{TeO2nonscintilla}
I.~Dafinei et. al: \textit{Nuclear Instruments and Methods in Physics Research} \textbf{A554}, (2005) 195-200; I.~Dafinei et al., \textit{Physica Status Solidi (A)} \textbf{204}, (2007) 1567-1570

\bibitem{Cerenkov}
T.~Tabarelli de Fatis: \textit{European Physical Journal C} \textbf{65}, (2010) 359-361

\bibitem{reflective_foil}
M.F.~Weber et al.: \textit{Science}  \textbf{287} (2000) 2451-2456

\bibitem{reflective_foil_scintillates}
R.~Lang et al.:  \textit{Astroparticle Physics} \textbf{33} (2010) 60-64

\bibitem{scintillating_foil_at_lowT}
 B.~van~den~Brandt et al: \textit{Nuclear Instruments and Methods in Physics Research}  \textbf{A446} (2000) 592-599; B.~van~den~Brandt, private communication

\bibitem{plastic_scintillator_at_lowE}
F.D~Becchetti et al.: \textit{Nuclear Instruments and Methods in Physics Research}  \textbf{A138} (1976) 93-104;  S.K.~Saraf et al.: \textit{Nuclear Instruments and Methods in Physics Research}  \textbf{A288} (1990) 451-454

 \end{thebibliography}



 \end{document}